\documentclass[aps,prl,twocolumn,groupedaddress,showpacs,amsmath,amssymb,floatfix,superscriptaddress]{revtex4-1}
\usepackage{multirow}
\usepackage{graphicx}
\usepackage{times}
\usepackage{color}
\usepackage{natbib}
\usepackage{array}
\newcolumntype{C}[1]{>{\centering\let\newline\\\arraybackslash\hspace{0pt}}m{#1}}

\newcommand{\sneos}{$S_{\textrm{neos}}$}

\newcommand{\sdyb}{$S_{\textrm{dyb}}$}

\newcommand{\shmvs}{$S_{\textrm{hmv}}\,$}
\newcommand{\sdybs}{$S_{\textrm{dyb}}\,$}

\begin{document}

\title{Sterile Neutrino Search at the NEOS Experiment}

\newcommand{\kaeri}{\affiliation{Neutron Science Division, Korea Atomic Energy Research Institute, Daejeon 34057, Korea}}
\newcommand{\cupibs}{\affiliation{Center for Underground Physics,
    Institute for Basic Science (IBS), Daejeon 34047, Korea}}
\newcommand{\cnu}{\affiliation{Department of Physics, Chonnam National University, Gwangju 61186, Korea}}
\newcommand{\knu}{\affiliation{Department of Physics, Kyungpook National University, Daegu 41566, Korea}}
\newcommand{\sju}{\affiliation{Department of Physics and Astronomy, Sejong University, Seoul 05006, Korea}}
\newcommand{\cau}{\affiliation{Department of Physics, Chung-Ang University, Seoul 06974, Korea}}
\newcommand{\kriss}{\affiliation{Korea Research Institute of Standards
and Science, Daejeon 34113, Korea}}
\newcommand{\ust}{\affiliation{University of Science and Technology,
Daejeon 34113, Korea}}

\author{Y.J.~Ko}\cau
\author{B.R.~Kim}\cnu
\author{J.Y.~Kim}\sju
\author{B.Y.~Han}\kaeri
\author{C.H.~Jang}\cau
\author{E.J.~Jeon}\cupibs
\author{K.K.~Joo}\cnu
\author{H.J.~Kim}\knu
\author{H.S.~Kim}\sju
\author{Y.D.~Kim}\cupibs\sju\ust
\author{Jaison Lee}
 \email{Corresponding author : jsahnlee@ibs.re.kr}
 \cupibs
\author{J.Y.~Lee}\knu
\author{M.H.~Lee}\cupibs
\author{Y.M.~Oh}
 \email{Corresponding author : yoomin@ibs.re.kr}
 \cupibs
\author{H.K.~Park}\cupibs\ust
\author{H.S.~Park}\kriss
\author{K.S.~Park}\cupibs
\author{K.M.~Seo}\sju
\author{Kim Siyeon}\cau
\author{G.M.~Sun}\kaeri

\collaboration{NEOS Collaboration}\noaffiliation

\date{\today}

\begin{abstract}
An experiment to search for light sterile neutrinos is conducted at a reactor 
with a thermal power of  2.8~GW located at the Hanbit nuclear power
complex. The search is done with a detector consisting of a ton of
Gd-loaded liquid scintillator
in a tendon gallery approximately 24~m from the reactor core.
The measured antineutrino event rate is 1976 per day with a
signal to background ratio of about 22. The shape of the antineutrino energy spectrum
obtained from the eight-month data-taking period is compared with a hypothesis of
oscillations due to active-sterile antineutrino mixing. No strong
evidence of 3+1 neutrino oscillation is found. An excess around the 5~MeV prompt
energy range is observed as seen in existing longer-baseline
experiments. The mixing parameter $\sin^{2}2\theta_{14}$ is limited up to
less than 0.1 for $\Delta m^{2}_{41}$ ranging from
0.2 to 2.3~eV$^{2}$ with a 90\% confidence level.
\end{abstract}


\maketitle

The mixing among three neutrinos has been
well established by experiments performed in the past
two decades since the discovery of neutrino oscillations~\cite{Fukuda:1998mi,Ahmad:2002jz, Eguchi:2002dm}. Consistent measurements
of the two mass differences and the three mixing angles of the standard, three-neutrino
mixing model have been reported by oscillation experiments using atmospheric, solar, reactor, and
accelerator neutrinos \cite{Fogli:2012ua}. Nevertheless, the mass hierarchy, the mass of the lightest neutrino, the Dirac or
Majorana nature of the neutrino, and the $CP$ phase are yet to be
determined~\cite{Giunti:2007ry}.

Even though the number of active light neutrinos is limited to
three by $Z$ boson decay-width measurements~\cite{ALEPH:2005ab}, it is
still possible to have
additional neutrinos if they are sterile. 
Sterile neutrinos can be identified by the occurrence of
active-sterile neutrino oscillations. A hint for this is
the LSND experiment's report of an observation of $\bar{\nu}_{\mu} \rightarrow \bar{\nu}_{e}$
mixing with a frequency corresponding to a mass-squared difference larger than
0.01~eV$^{2}$~\cite{Athanassopoulos:1996jb}.
Results from the MiniBooNE's test of the LSND signal are, however, inconclusive~\cite{Aguilar-Arevalo:2013pmq}.

In addition to the LSND result, there are two other anomalies that could possibly be signs
of active-sterile neutrino oscillations. An apparent $\nu_e$ disappearance over a baseline
of a few meters in the GALLEX and SAGE  gallium experiments exposed to
radioactive sources was reported~\cite{Abdurashitov:2005tb}; the ratio of the
numbers of measured and predicted events is  $\rm 0.88 \pm 0.05$.
A number of short-baseline reactor antineutrino experiments
established limits on the presence of neutrino oscillations with eV
mass differences by shape analyses of the measured neutrino energy
spectra.
Among those experiments, the Bugey
experimental limits on sterile neutrinos are the most stringent~\cite{Declais:1994su}. 
Mueller {\em et al.}~\cite{Mueller:2011nm} found about a 6\% deficit in reactor antineutrino event rates
compared with the theoretical expectations for the short-baseline reactor
experiments, which is the so-called ``reactor antineutrino anomaly''
(RAA).
It can be interpreted as an active-sterile
neutrino oscillation with three active neutrinos plus one or more sterile neutrinos, i.e., a
$3+n~\nu$ scenario~\cite{Mention:2011rk,Huber:2011wv}, compatible with the LSND result.
Recent reactor experiments that measured the $\theta_{13}$ mixing angle, Daya
Bay \cite{An:2015nua}, RENO \cite{RENO:2015ksa}, and Double Chooz \cite{Abe:2011fz}, all confirmed
a similar deficit in the measured neutrino event rates. It is also
intriguing that these three experiments observed a significant event
excess beyond expectations from existing reactor-flux
models~\cite{Huber:2011wv, Dwyer:2014eka, Hayes:2013wra} at prompt
energies around 5~MeV.

The Planck satellite experiment~\cite{Ade:2015xua} constrained the
effective neutrino number to less than 3.7 at a 95\% confidence level
and excluded the existence of sterile neutrinos with masses near 1 eV
fully thermalized in the early Universe. However, theoretical models such as a large lepton asymmetry~\cite{Hamann:2010bk} or neutrino self-interactions~\cite{Hannestad:2013ana} show that the effective
number of sterile neutrinos can be much less than one. Therefore,
light sterile neutrinos remain compatible with current cosmological
constraints and should be searched for in more refined experiments
with higher sensitivities. The phenomenology of light sterile
neutrinos was recently reviewed in Ref.~\cite{Gariazzo:2015rra}.

A search for sterile neutrinos at a nuclear reactor was first proposed by
Mikaelyan \cite{Mikaelyan:1998yg}. Following the 3+1 $\nu$ mixing scenario
\cite{Giunti:2011gz}, the survival probability of a neutrino
with energy $E_{\nu}$ at a distance $L$
shorter than 100~m can be approximated as
\begin{equation}
P \simeq 1-\sin^{2}2\theta_{14}\sin^{2}\left[ 1.27\frac{\Delta
m^{2}_{41}L}{E_{\nu}} \left( \frac{\textrm{eV}^{2}\cdot\textrm{m}}{\textrm{MeV}}\right)\right]\,.
\label{eq:surv_prob}
\end{equation}
A new oscillation parameter set of
($\sin^{2}2\theta_{14},\,\Delta m^{2}_{41}$) introduced by
the existence of an eV-scale light sterile neutrino can be obtained by
measuring the distortion in the energy spectrum and/or a deficit from
the expected number of inverse beta decay (IBD, $\bar{\nu}_{e}+p\rightarrow
e^{+}+n$) events at a short
distance from a nuclear reactor core.
Considering the IBD energy spectrum, which is smoothly
peaked at around 3 MeV, the sensitivity for observing an $\sim$1 eV$^{2}$
mass-squared difference becomes the highest at several meters and falls off as
the distance increases. The Daya Bay experiment sets limits on
a light sterile neutrino with lower (i.e., $<1$~eV$^2$) mass-squared differences~\cite{An:2016luf}.
Currently, a number of short-baseline reactor experiments are being developed~\cite{Vogel:2015wua}.
Here we report results from the NEOS (neutrino experiment for oscillation at short baseline)
experiment for a light sterile neutrino search at a distance of 24~m
from a reactor core.

The NEOS detector was installed in the tendon gallery of reactor
unit 5 of the Hanbit Nuclear Power Complex in Yeonggwang, Korea.
This is the same reactor complex being used for the RENO
experiment~\cite{RENO:2015ksa}.  The active core size of unit 5
is 3.1~m in diameter and 3.8~m in height and contains 177 low-enriched
uranium fuel assemblies; about one-third of these assemblies are
replaced with fresh
ones every 18 months. The tendon gallery is located 10~m
below ground level and is directly under the wall of the containment
building. The minimum overburden with the ground and building
structures corresponds to 20~m water equivalent. The detector
is centered at 23.7$\pm$0.3~m from the center of the reactor core,
while the distance to the closest neighboring reactor core is 256~m. 

\begin{figure}
\centering
\includegraphics[width=0.46\textwidth]{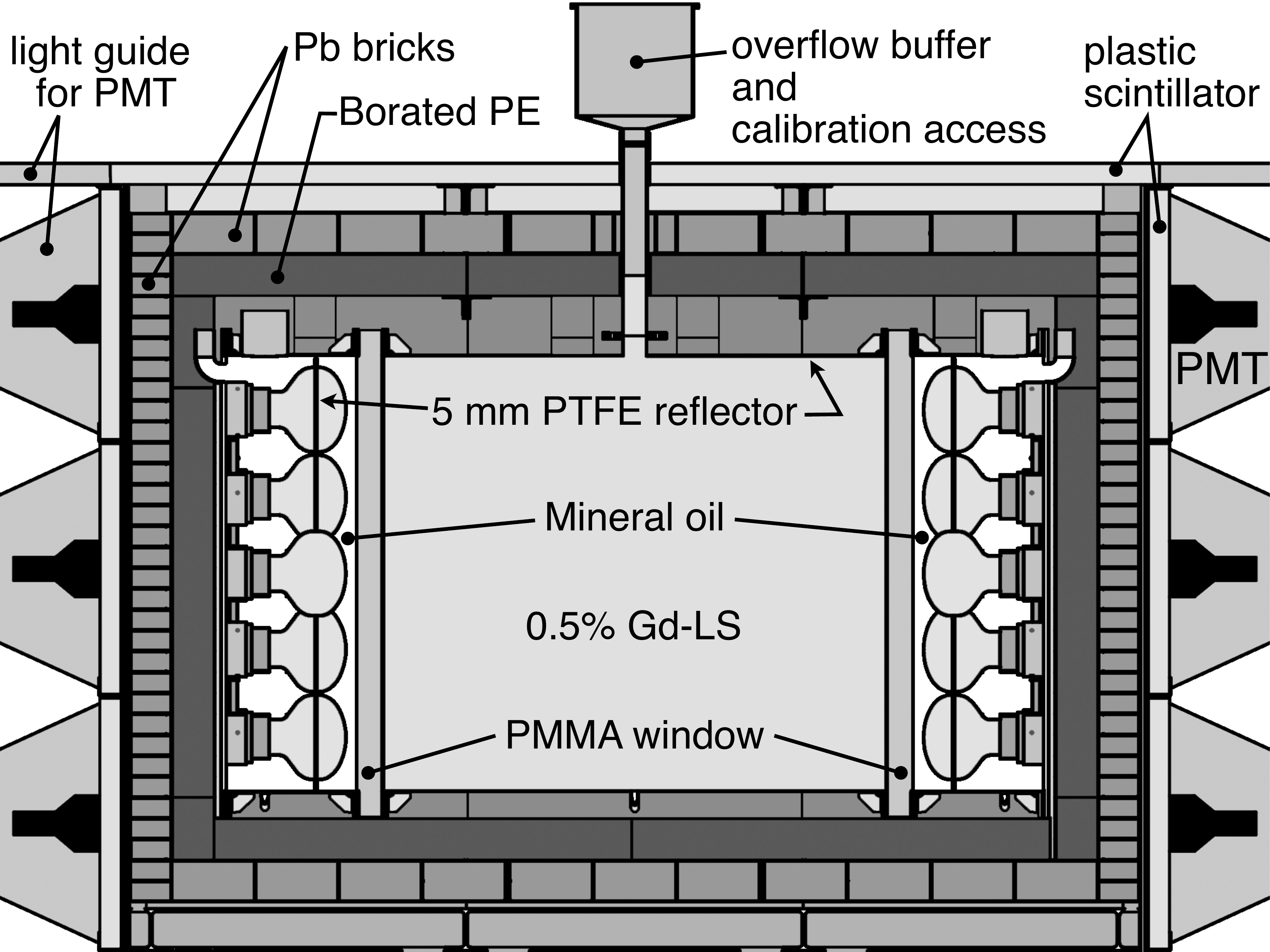}
\caption{A simplified cross-sectional view of the NEOS
detector.}
\label{fig:detector}
\end{figure}

The NEOS detector consists of a neutrino target, mineral oil buffers,
passive shieldings, muon counters, and supporting structures (see Fig.~\ref{fig:detector}).
The positron annihilation followed by a neutron capture from an electron
antineutrino IBD process is detected in the target, which
is a horizontal cylindrical stainless-steel tank with a 1008~L inner
volume (103~cm in diameter and 121~cm in length)
filled with a 0.5\%~Gd-doped liquid scintillator~\cite{Kim:2015qlu}.  Each end of the 
target vessel is viewed by 19 eight-inch photomultiplier tubes (PMTs) that are closely packed
in mineral oil buffers. Each buffer and target are separated
by a 6-cm-thick transparent PMMA window. Plates of 5-mm-thick
polytetrafluoroethylene (PTFE)
reflector are installed on the inner wall of the target vessel and along
the PMT glasses' equator surfaces.
The target tank is enclosed by a 10-cm-thick- borated PE and lead layers 
for shielding neutrons and external gamma rays, respectively. Muon counters made from 5-cm thick plastic scintillators
surround the outside of the detector.

The waveforms of all 38 PMTs are digitized and recorded by
500 megasampling (MS) per second flash
analog-to-digital converter (ADC) modules, each of which makes an independent trigger decision. Signals from the
muon counters are processed by a 62.5 MS/s ADC module. A trigger
control board decides
the global trigger and synchronizes the ADC modules. A trigger
requiring 30 or more PMT signals higher than the 6~mV threshold is fully
efficient for energies above 400~keV.
The trigger rate was about 210 Hz.
The detector operated for 46 days with the reactor off ($t_{\textrm{off}}$) and 180 days
with the reactor on ($t_{\textrm{on}}$).

\begin{figure}[b]
\centering
\includegraphics[width=0.48\textwidth]{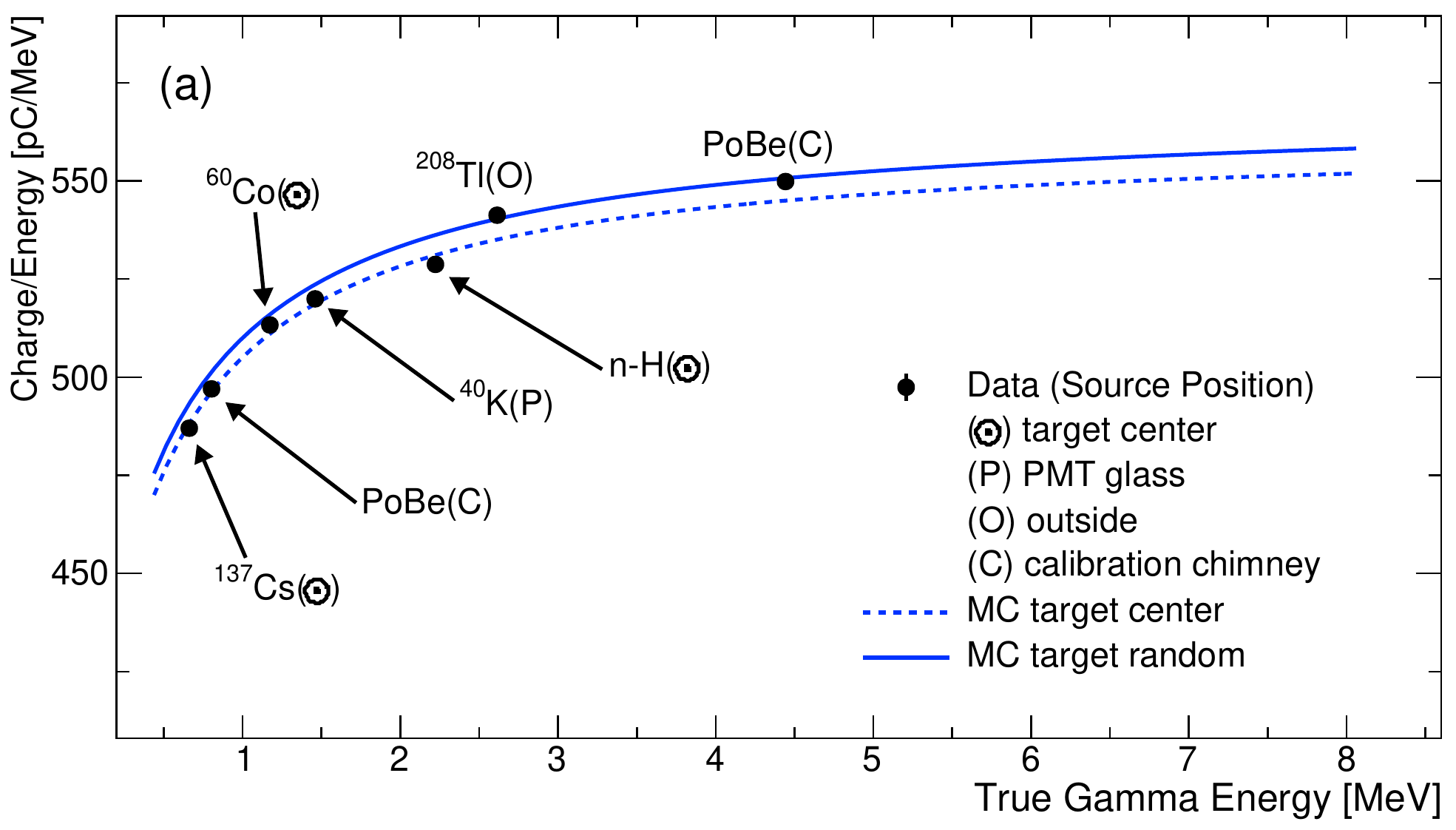}\\
\includegraphics[width=0.48\textwidth]{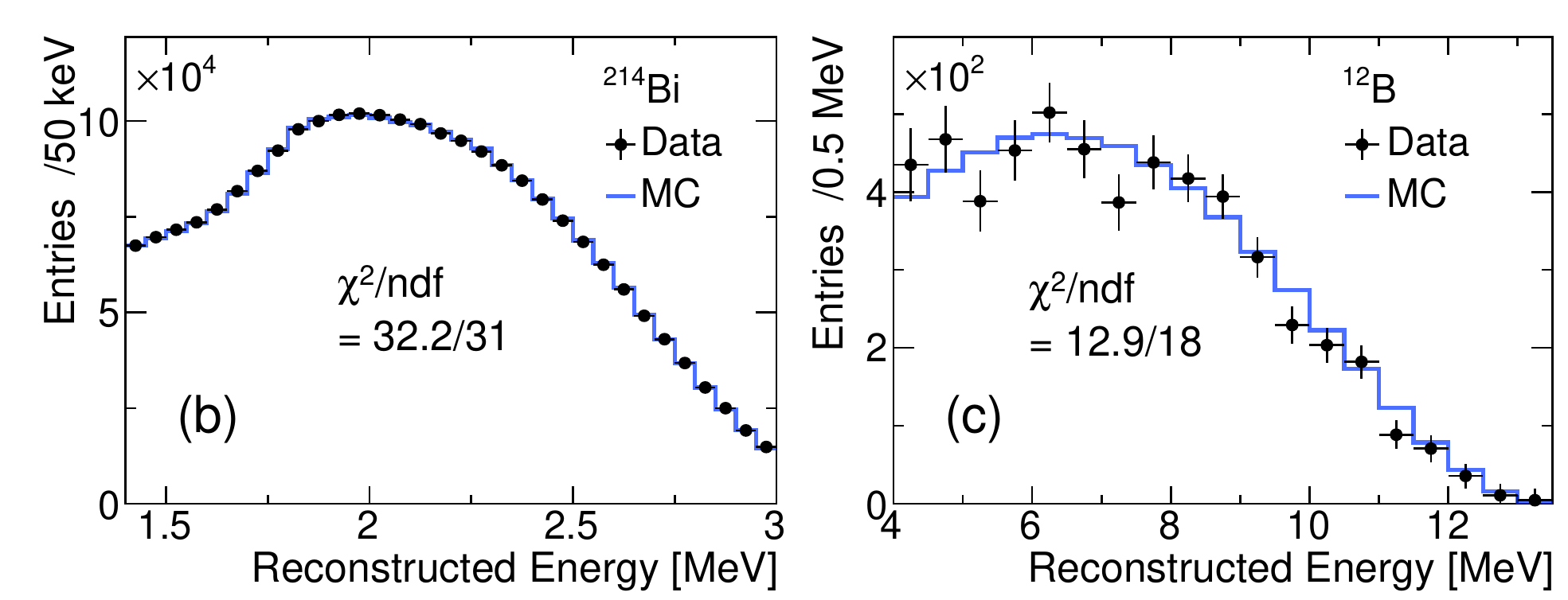}
\caption{Detector responses to $\gamma$ and $\beta$ sources: 
(a) ratios of full peak charges to the true $\gamma$ energies, (b)
  $\beta$-decay spectra for $^{214}$Bi, and (c) $^{12}$B.}\label{fig:eRecon}
\end{figure}
The detector was calibrated once every week with $^{137}$Cs,
$^{60}$Co, $^{252}$Cf, and PoBe sources. Continuous
background events from several well-known radioactivities are used for
additional calibrations.
The charge to energy ratios of single gamma ray events show a nonlinear 
detector response as shown in Fig.~\ref{fig:eRecon}(a). 
An empirical function used to describe this nonlinearity 
is
\begin{equation}
Q/E_{\gamma}=\left(p_{0}+p_{1}E_{\gamma}\right)\lbrack
1+p_{2}\textrm{exp}(p_{3}\sqrt{E_{\gamma}})\rbrack,
\label{eq:qe_nonlin}
\end{equation}
where $Q$ is the charge, $E_{\gamma}$ is  the true
$\gamma$ energy, and $p_{i}$ terms are fitting parameters.
The detector stability and the nonuniform response along the horizontal 
axis of the detector are continuously monitored and corrected using
2.6~MeV external $\gamma$ rays from $^{208}$Tl and internal $\alpha$
background events.

The detector is simulated with a {\sc geant}4-based Monte Carlo (MC)
simulation~\cite{Agostinelli:2002hh}. The optical properties of the liquid scintillator and reflecting
materials and responses of PMTs and electronics are fine-tuned to describe the
source calibration data, and, consequently, the effects of
escaping $\gamma$ rays, energy resolution
($\sigma/E_{\gamma} \sim 5$\% for a full peak at 1~MeV), and the nonlinear $Q$ to
$E_{\gamma}$ response are well reproduced.
The reconstructed energy spectra for  $^{214}$Bi and $^{12}$B $\beta$ decays are
shown in Figs.~\ref{fig:eRecon}(b) and \ref{fig:eRecon}(c) with the MC results
superimposed. The systematic error on the energy scale associated with 
differences between the data and MC calculations is 0.5\%.

The selection criteria of IBD candidate events are determined to maximize the signal to
background ratio. We start with a pair of events which consists of a prompt event candidate that has
an energy above 1 MeV and its following delayed event candidate of an $n$-Gd
capture signal with energy between 4 and 10 MeV in a 1--30~$\mu$s time
window. To
exclude multiple neutron-induced backgrounds, the pair is rejected when any
events occur at a time that is less than 30~$\mu$s before or 150~$\mu$s after
the prompt signal time. Pairs of which the prompt or delayed events
occur in a 150~$\mu$s interval after a muon-counter hit are vetoed. 
Finally, pairs caused by the scattering and subsequent capture of
fast neutrons are identified using a pulse shape discrimination (PSD) requirement
that is adjusted to accept more than 99.9\% of the electron-induced recoil events
over the full energy range. The background fraction that is removed by the PSD
requirement was measured to be 73\% during the reactor-off period.

With these requirements, $1976.7\pm3.4$~($85.1\pm1.4$) IBD candidates
per day were selected during the reactor-on~(-off) period with the prompt energy between 1 and 10~MeV.
No evidence was found for
additional backgrounds associated with the reactor operation or for
significant background fluctuation in the whole running period.
The muon-counter rate, to which the fast-neutron
background is related, was stable at 241~Hz with a 2~Hz day-to-day rms variation,
The energy distributions of the fast-neutron scattering events that
were rejected by the PSD requirement show only small variations consistent
with statistical fluctuations throughout the entire running period.
Contributions from accidental background events were estimated by the time-delayed
coincidences method~\cite{Abe:2012tg} to be $7\pm1$ per day, where the
error corresponds to the range of the daily variations.

\begin{figure}[!t]
\centering
\includegraphics[width=0.48\textwidth]{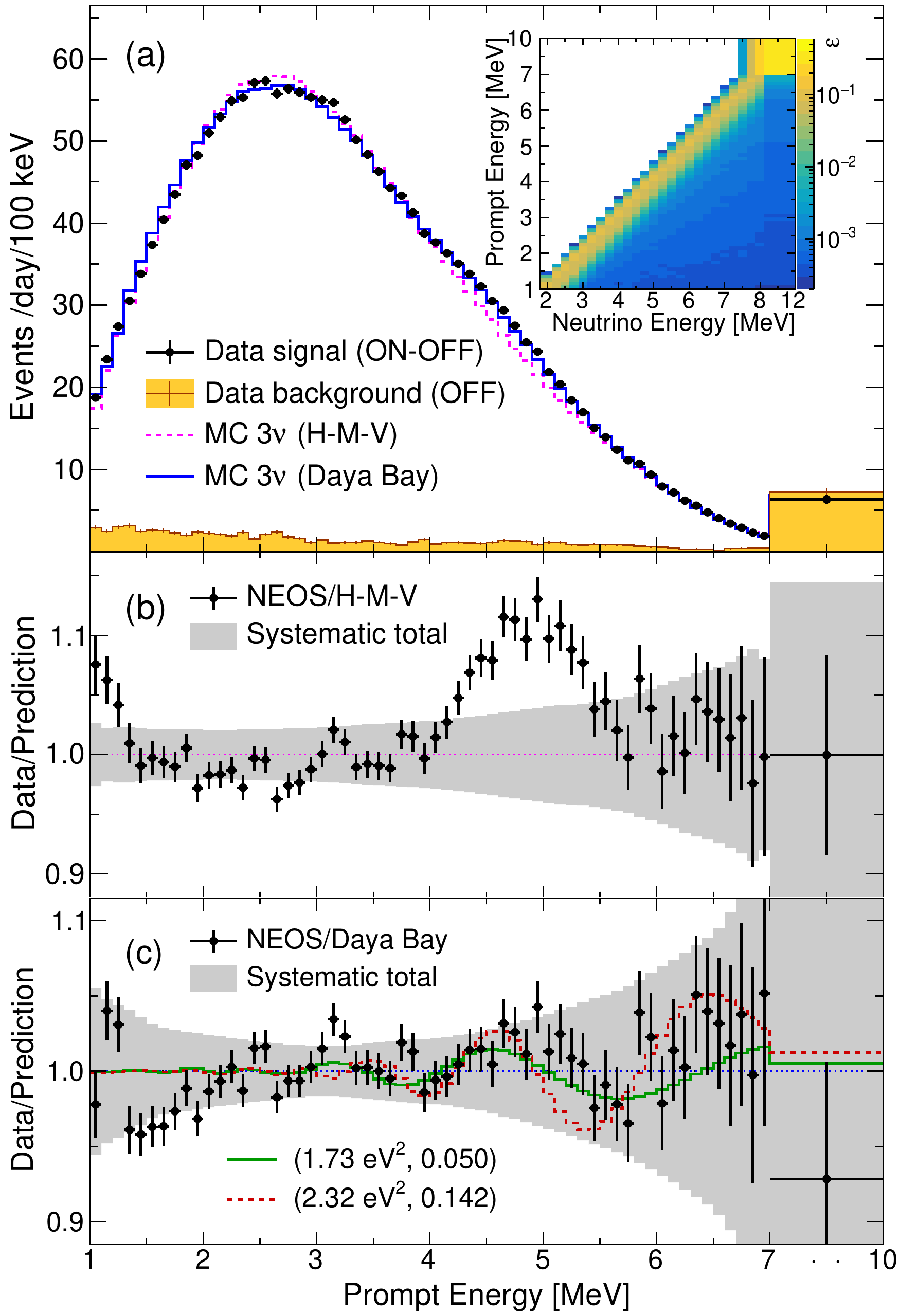}
\caption{(a) The IBD prompt energy spectrum. The last bin is
  integrated up to 10~MeV. The orange shaded histogram is the background spectrum measured during the
  reactor-off period. The detector response matrix in the inset shows
  the relation between the neutrino energy and the prompt energy. (b) The
  ratio of the observed prompt energy spectrum to the HM flux prediction
  weighted by the IBD cross section with the 3$\nu$ hypothesis. The predicted
  spectrum is scaled to match the area of the data excluding the 5~MeV
  excess region (3.4--6.3~MeV).
(c) The ratio of the data to the expected
spectrum based on the Daya Bay result with the 3$\nu$ hypothesis, scaled
to match the whole data area.
The solid green line is the expected oscillation patterns for the
best fit of the data to the 3+1 $\nu$ hypothesis and the corresponding
oscillation parameters $(\sin^{2}2\theta_{14},\,\Delta m^{2}_{41})$ is
(0.05, 1.73~eV$^{2}$). The dashed red line is the expected oscillation pattern for the RAA best fit
parameters (0.142, 2.32~eV$^{2}$). The gray error bands in (b) and (c)
are estimated total systematic uncertainties,
corresponding to the square roots of diagonal elements of the covariance
matrices.}
\label{fig:spectrum}
\end{figure}

The measured prompt energy spectrum (\sneos) is shown in
Fig.~\ref{fig:spectrum}(a), superimposed with the predicted
nonoscillation spectra:
one based on flux calculations by Huber~\cite{Huber:2011wv} and
Mueller (HM)~\cite{Mueller:2011nm}
weighted by the IBD cross sections estimated by Vogel and Beacom~\cite{Vogel:1999zy}, and 
another based on the Daya Bay reactor antineutrino
spectrum~\cite{An:2016srz}. The former and the latter predicted
spectra are denoted as \shmvs and \sdyb, respectively, and their
superscript 3$\nu$~(4$\nu$) denotes the 3~(3+1) $\nu$
hypothesis. The predicted spectra are
generated using the detector response shown in the inset in
Fig.~\ref{fig:spectrum}(a) produced by a full simulation of IBD events of which the
$\bar{\nu}_{e}+p$ reaction
occurs at random positions throughout the detector target and produced $e^{+}$ and $n$ are propagated through all of
the detector responses. The antineutrinos are assumed to originate
uniformly throughout the cylindrical active reactor
core and the average primary element fission fractions of 0.655, 0.072, 0.235, and 0.038 for 
$^{235}$U, $^{238}$U, $^{239}$Pu, and $^{241}$Pu, respectively, are
used. The differences between the fission fractions for the NEOS data and
the ones for Daya Bay are taken into account and small corrections are
made using the HM flux model as instructed in Ref.~\cite{An:2016srz}.

The excess around 5~MeV versus $S_{\textrm{hmv}}^{3\nu}$ is clearly seen, as
shown in Figs.~\ref{fig:spectrum}(a) and \ref{fig:spectrum}(b), for
the first time at this short baseline whereas previous short-baseline
measurements~\cite{Kwon:1981ua,Declais:1994su} did not show a
clear excess. The excess does not completely
disappear even when the data are divided by $S_{\textrm{dyb}}^{3\nu}$, as
shown in
Fig.~\ref{fig:spectrum}(c). This can be explained as that the
excess can be contributed differently from each fission
element~\cite{Huber:2016xis}. It is, however, difficult to conclude
with the current level of uncertainties.
Another large discrepancy other than the 5~MeV excess for the $S_{\textrm{hmv}}^{3\nu}$~ case is found at the lowest
energy range. The disagreement is as large as 8\% at 1~MeV and drops rapidly as
the energy increases. For the incident antineutrino flux below 2~MeV and
above 8 MeV, where
tabulated data do not exist, we used the
exponential functions in Refs.~\cite{Huber:2016xis,Mueller:2011nm} for
an extrapolation. For the comparison with \sdyb, the fluctuation shown
in the lowest energy range is mainly due to the convolution of the
spectrum from the original one with large neutrino energy bin sizes to
one with finer prompt energy bin sizes for
this work. Other small fluctuations at
several energies also seem to have some small structures which are
common for both reference spectra but, regarding the
uncertainties, are not so
significant.

The following systematic uncertainties are taken into account.
Errors in the reference antineutrino spectra are the main
contributors to the total uncertainties. The 0.5\% uncertainty in
the reconstructed energy scale is another large contributor to
the total uncertainty.
Other sources of uncertainty, such as the inaccuracy of the effective
baseline, fuel-related uncertainties from burn-up and fission fractions, spill-in from inactive volumes, events generated by antineutrinos from neighbor reactors, and
other detector-related uncertainties have negligible effects on the
spectral shape.


\begin{figure}[b]
\centering
\includegraphics[width=0.48\textwidth]{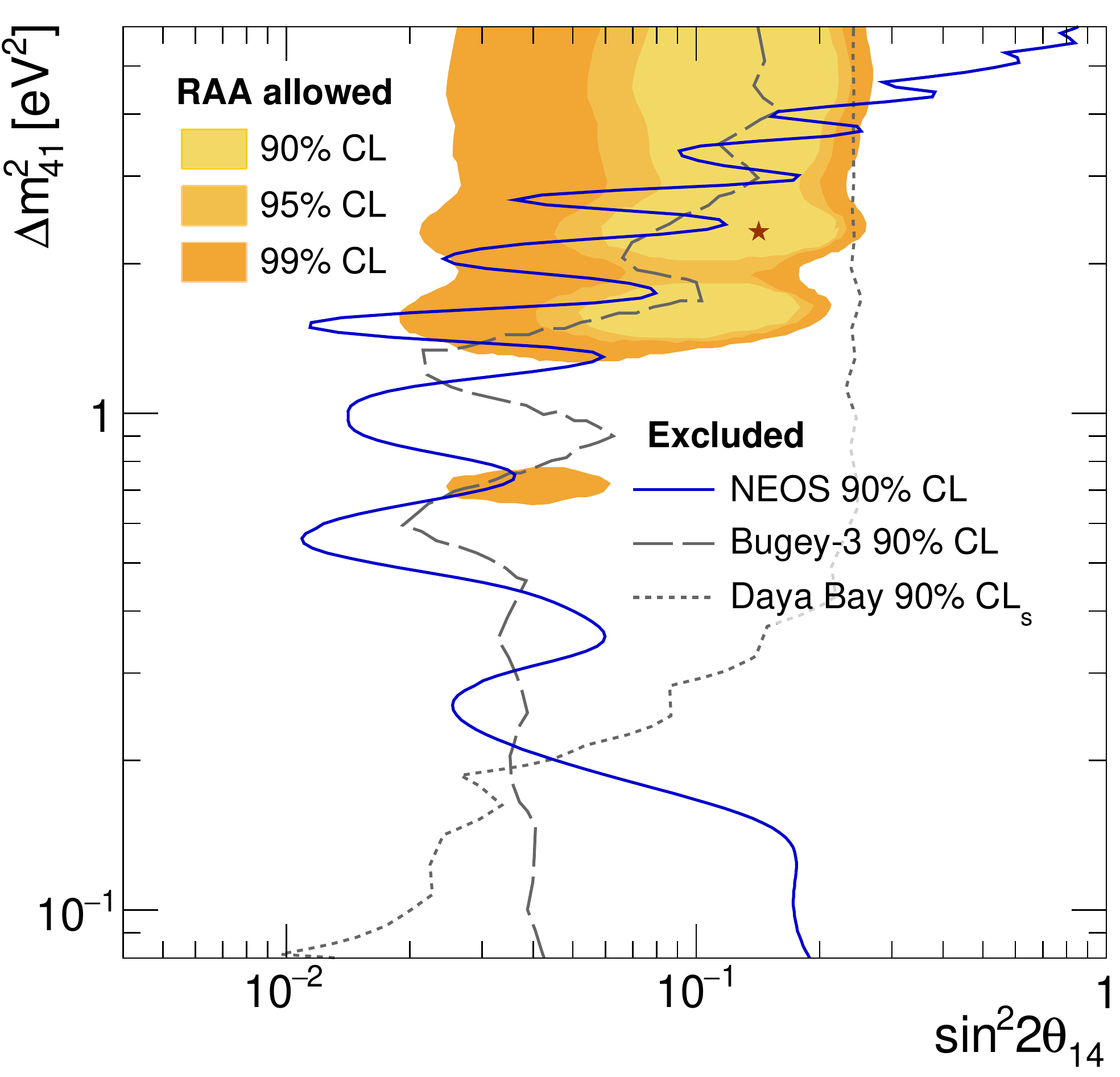}
\caption{Exclusion curves for 3+1 neutrino oscillations in the $\sin^22\theta_{14}-\Delta m_{41}^2$
parameter space. The solid blue curve is 90\% C.L. exclusion contours based
on the comparison with the Daya Bay spectrum, and the dashed gray curve is
the Bugey-3 90\% C.L. result~\cite{Declais:1994su}. 
The dotted curve shows the Daya Bay 90\% CL$_{\textrm{s}}$ result~\cite{Adamson:2016jku}.
The shaded area is the allowed region from the reactor antineutrino anomaly fit,
and the star is its optimum point~\cite{Mention:2011rk}.}
\label{fig:exclude}
\end{figure}

Probing an oscillation in a spectrum measured with a single
detector at one fixed distance from the reactor core depends on the
accuracy and precision of the reference spectrum.
Among the available references, the flux calculation by Huber and Mueller
provides tabulated uncertainties with their correlations between the
neutrino energy bins and isotopes and,
even though their uncertainties are underestimated~\cite{Hayes:2013wra},
their spectral shapes (not their absolute rates)
are generally in good agreement with existing experimental results except for the
region of the 5 MeV excess. A recent
high-resolution {\em ab initio} calculation by Dwyer and Langford~\cite{Dwyer:2014eka} better describes the observed
5~MeV excess, but its large uncertainties and their
correlations, which are yet to be exactly quantified, make a comparison
with our data impractical. Experimentally, only the Daya Bay unfolded
spectrum~\cite{An:2016srz} is based on a direct measurement and, therefore, the
uncertainties in the antineutrino spectrum are relatively small. The
correlation of uncertainties among the energy bins can be dealt with
by the provided covariance matrix.

In the present work, the measured prompt energy spectrum is compared
with \sdybs for
testing the oscillation.
A $\chi^{2}$ is constructed with 61 data points in 1--10~MeV prompt
energy spectrum and a covariance matrix $V_{ij}$
that accounts for correlations between uncertainties:
\begin{equation}
\chi^{2} = \sum_{i=1}^{N}\sum_{j=1}^{N}(M_{i}-\frac{t_{\textrm{on}}}{t_{\textrm{off}}}B_{i}-T_{i})V_{ij}^{-1}(M_{j}-\frac{t_{\textrm{on}}}{t_{\textrm{off}}}B_{j}-T_{j})\,,
\end{equation}
where $M$~($B$) is the number of measured IBD candidate events accumulated
during the reactor-on (-off) period,
$T$ is the prediction from a reference spectrum that
accounts for oscillation parameters, and the subscripts $i$ and $j$ denote the
prompt energy bin.
To construct $V_{ij}$, the elements for the errors in the reference antineutrino spectrum are
calculated from the matrix in Table 13 of
Ref.~\cite{An:2016srz}, by convolving them with the detector
response shown in the inset in Fig.~\ref{fig:spectrum}(a).
Then the other elements from statistical and detector systematic
uncertainties are added.

The $\chi^{2}$ values are calculated on a fine grid in the sensitive $\Delta
m^{2}_{41}$ range from 0.06 to 6~eV$^{2}$.
The $\chi^{2}$ value with the 3$\nu$
hypothesis is $\chi^{2}_{3\nu}/\textrm{NDF}=64.0/61$, where NDF denotes the number of degrees of freedom. The minimum $\chi^{2}$ value
with the 3+1 $\nu$ hypothesis, $\chi^{2}_{4\nu}/\textrm{NDF}=57.5/59$, is
obtained at $(\sin^{2}2\theta_{14},\,\Delta m^{2}_{41})=(0.05,
1.73~\textrm{eV}^{2})$, and the second minimum at $(0.04, 1.30~\textrm{eV}^{2})$
has a similar $\chi^{2}$ value to the first one. The values of the mass-squared
differences of the two minima are compatible with the latest
global fit results~\cite{Kopp:2013vaa,Gariazzo:2015rra}, though the mixing angle
parameters, $\sin^{2}2\theta_{14}$, are smaller than
those global best fit values.
The $p$ value corresponding to the
$\chi^{2}$ difference between the 3$\nu$ hypothesis and the best fit
for the 3+1 $\nu$ hypothesis, $\Delta
\chi^{2}=\chi^{2}_{3\nu}-\chi^{2}_{4\nu}=6.5$, is estimated to be 22\%
using a large number of Monte Carlo data sets with statistical and
systematic fluctuations~\cite{Fig:supple}.
As a result, no apparent parameter set of
$(\sin^{2}2\theta_{14},\,\Delta m^{2}_{41})$ that has significant
favor for the 3+1 $\nu$ hypothesis is found.

The limit on the $\sin^{2}2\theta_{14}$ value for each $\Delta
m^{2}_{41}$ is found using a raster scan~\cite{Lyons:2014kta}.
For a $\Delta m^{2}_{41}$ value, a probability density function
$f(\sin^{2}2\theta_{14})$ is
constructed from the $\Delta \chi^{2}$ distribution in
the $\sin^{2}2\theta_{14}$ range from 0 to 1, where $\Delta \chi^{2}$
is the difference between a $\chi^{2}$ value at a $\sin^{2}2\theta_{14}$ point
and the minimum $\chi^{2}$ value at the corresponding $\Delta
m^{2}_{41}$. The upper limit
($ul$) at
confidence level (C.L.) of $1-\alpha$ is found with the condition of
\begin{equation}
\int_{ul}^{1}
f(\sin^{2}2\theta)d(\sin^{2}2\theta)= \alpha.
\label{eq:cl}
\end{equation}
The resulting exclusion limits at 90\% C.L. are shown in
Fig.~\ref{fig:exclude}, superimposed with the 90\% C.L.
exclusion curves of the Bugey-3~\cite{Declais:1994su} and the Daya Bay~\cite{Adamson:2016jku} limits and with the allowed
region by the RAA fit from Fig.~8 in
Ref.~\cite{Mention:2011rk}. The mixing angle parameter
$\sin^{2}2\theta_{14}$ is excluded for the region significantly less
than 0.1 for the
$0.2~\textrm{eV}^{2}<\Delta m^{2}_{41}<2.3~\textrm{eV}^{2}$ range. Our limits are compatible with the Bugey-3
result at the 0.2~eV$^{2}< \Delta m^{2}_{41}<$ 4~eV$^{2}$ range, since the 
baselines of the two experiments are similar. The Bugey-3 and the Daya Bay
are more sensitive than the NEOS at lower $\Delta
m^{2}_{41}$ because of their spans of longer baselines. At above 4~eV$^{2}$, our
sensitivity drops as a natural consequence of shape-only analysis, while the Bugey-3
and the Daya Bay limits converge to constant $\sin^{2}2\theta_{14}$ values since the absolute rates are taken
into account based on the
ILL-Vogel~\cite{Schreckenbach:1985ep,Vogel:1983hi} and Huber-Mueller flux models in their
analyses, respectively.
Our limit curve shows a more ragged shape than that of
Bugey-3, because the differences between the data and the
model spectrum are more significant by higher statistics and by use
of the Daya Bay model spectrum which has smaller errors around the
spectral peak range. For a more practical comparison and/or combined
analysis of this work with the Bugey-3 data, it would be necessary to
revise the Bugey-3 data with the Daya Bay absolute spectrum, which
should be more realistic for measurements using similar types of commercial reactors.

In conclusion, no strong evidence for 3+1 neutrino oscillations is observed in
this study. We could set up new stringent upper limits on the $\theta_{14}$ mixing
angle for the $\Delta m^{2}_{41}\sim1~\textrm{eV}^{2}$ region, thanks
to the high signal to background ratio, good energy resolution,
and using the most realistic reference antineutrino spectrum.
The results are currently limited by uncertainties in the reference
spectrum of the Daya Bay and systematics of the NEOS
data. The systematic uncertainties in the antineutrino spectrum will
be reduced if the reference spectrum from the RENO experiment is available
since it
uses the same reactor complex. Other ongoing or scheduled
experiments~\cite{Ryder:2015sma,Ashenfelter:2013oaa,Helaine:2016bmc,Alekseev:2016llm}
with even shorter
baselines and/or better $L/E$ resolution are expected to improve the sensitivities. It should be remarked that, in addition to these  
short-baseline sterile neutrino searches, future long-baseline reactor antineutrino
experiments~\cite{An:2015jdp,Kim:2014rfa} aimed at the determination
of the neutrino mass hierarchy would require more accurate reference
IBD spectra. Recently, the IceCube and MINOS experiments constrained
the $\theta_{24}$ mixing angle forthe  3+1 $\nu$
model and
rejected the LSND anomaly parameter space for $\Delta m^{2}_{41} <
3$~eV$^{2}$~\cite{TheIceCube:2016oqi,MINOS:2016viw}, for which the former assumed $\theta_{14}$ values from
the global fits~\cite{Kopp:2013vaa,Conrad:2012qt} and the latter combined
$\theta_{14}$ constraints from the Daya Bay and the Bugey-3 results~\cite{Adamson:2016jku}. Our new limit will further improve the constraints to the LSND anomaly parameter space by combining with the $\theta_{24}$ measurements.

This work is supported by IBS-R016-D1 and 2012M2B2A6029111 from
National Research Foundation (NRF).
We appreciate the  Korea Hydro and Nuclear Power
(KHNP) company and especially acknowledge the help and support
provided by staff members of the
Safety and Engineering Support Team of Hanbit Nuclear Power Plant 3. We appreciate the Daya Bay
Collaboration for the discussion on the errors in their reference spectrum.

\bibliography{sterileneutrinos}

\end{document}